# Photonic Neuromorphic Accelerator for Convolutional Neural Networks based on an Integrated Reconfigurable Mesh

ARIS TSIRIGOTIS,[1] GEORGE SARANTOGLOU,[1] STAVROS DELIGIANNIDIS[2], ERICA SÁNCHEZ[3], ANA GUTIERREZ[3], ADONIS BOGRIS,[2] JOSE CAPMANY[4] AND CHARIS MESARITAKIS[1,*]

[1]Dept. Information and Communication Systems Engineering, University of the Aegean, Palama 2 str. Karlovasi Samos 83200-Greece
[2]Dept. Informatics and Computer Engineering, University of West Attica, Ag. Spiridonos, 12243, Egaleo, Greece
[3]iPronics, Programmable Photonics, Valencia, Spain.
[4] Photonics Research Labs, Universitat Politècnica de València, Camino de Vera, 46022, Valencia, Spain
*cmesar@aegean.gr

**Abstract:** In this work, we present and experimentally validate a passive photonic-integrated neuromorphic accelerator that uses a hardware-friendly optical spectrum slicing technique through a reconfigurable silicon photonic mesh. The proposed scheme acts as an analogue convolutional engine, enabling information preprocessing in the optical domain, dimensionality reduction and extraction of spatio-temporal features. Numerical results demonstrate that utilizing only 7 passive photonic nodes, critical modules of a digital convolutional neural network can be replaced. As a result, a 98.6% accuracy on the MNIST dataset was achieved, with a power consumption reduction of at least 26% compared to digital CNNs. Experimental results confirm these findings, achieving 97.7% accuracy with only 3 passive nodes.

## 1. Introduction

The exploding growth of the Internet of everything (IoE) ecosystem [1] has unleashed the generation of a tremendous amount of raw data, demanding high-speed low-power processing. In this landscape, typical von-Neumann computers, characterized by their processing architecture, have met an efficiency road-block [2], struggling with data bottlenecks and energy constraints. Consequently, bio-inspired computing has emerged as an unconventional and promising route, aiming to circumvent inherent limitations of traditional systems [3]. This approach, drawing on the complex mechanisms of biological organisms, offers parallel processing capabilities and adaptability, potentially transforming the way we handle the escalating data challenges in the IoE era. Towards this route, photonic integrated circuits (PIC) can offer a proliferating hardware platform for neural network implementation, based on merits such as multiplexing assisted parallelism, low power consumption, large-scale integration and low latency [4–6].

One of the most common types of neural networks used for processing IoE generated visual data are convolutional neural networks (CNNs) [7]. Highly acclaimed for their ability to extract features from large datasets in a hierarchical manner, CNNs owe much of their effectiveness to their unique distinctive layered structure. Typically, CNN architectures consist of three types of layers: convolutional, pooling and fully-connected layers. The convolutional layers, functioning as the cornerstone of the network, apply multiple sets of weights (kernel filters) to the input for feature detection, transforming the data into a feature map. Subsequently, the pooling layers reduce the dimensionality of the feature maps by computing the maximum or the mean of a local patch of units in one or several feature maps. Finally, the fully connected layers integrate the processed features, performing high-level reasoning to facilitate the final decision-making or classification task.

A significant issue of CNNs is the high computational/power demand of convolutional operations, often consuming up to 90% of the network's execution time [8]. This has led to a focus on optimizing these operations in various hardware architectures, including traditional GPUs [9], memristors [10] and lately photonics [11]. These technologies primarily aim to enhance the efficiency of digital matrix-to-vector multiplications (MVMs), which are fundamental to executing convolutional operations in CNNs. In this context, integrated solutions such as micro-ring resonator (MRR) banks [12], photonic tensor cores that exploit frequency combs to encode the inputs or the kernel weights [13,14] and cascaded fixed Mach-Zehnder interferometers (MZI) [15] have risen as low-power MVM accelerators able to reach 1.27 Tera-operations per joule.

The adaptation of MVMs into the photonic domain introduces significant challenges. One primary issue is the necessity for extensive offline numerical simulations during high-speed weight adaptation in training. These simulations are meant to emulate the behavior of photonic hardware but often lead to power inefficiency and make photonic MVMs prone to errors in weights. These errors arise from discrepancies between the actual physical parameters of the chip and those assumed in the simulations. Additionally, photonic accelerators that implement MVMs typically have fixed dimensions and connectivity, tailored for specific tasks. This design choice limits their flexibility in adapting to new tasks and reduces the potential for chip reusability. Finally, the current scalability constraints of photonic platforms limit the number of total MVMs that can be executed in a single pass. Consequently, as the resolution and/or size of visual data increase, implementing multiple large size kernels necessitates several processing passes on the same PIC. This approach not only increases the total processing time but also elevates energy consumption.

In [16], the operational principles of an unconventional integrated photonic accelerator were presented. The proposed system, namely OSS-CNN, is comprised of two main components: an analog pre-processor, which utilizes high-speed passive photonic and opto-electronic devices for feature extraction, nonlinear transformation and dimensionality reduction and a digital post-processor, in which a simple feedforward neural network (FNN) is employed to associate the analog outputs with their corresponding class labels. The novelty of this acceleration approach stems from the adoption of an optical spectrum slicing (OSS) technique. This technique employs parallel passive optical filter nodes during the convolutional stage, significantly reducing the overall power consumption of a CNN system, as will be detailed in the follows sections. Moreover, this method introduces unparalled scalability by enabling convolution operations of any size, thus liberating the photonic hardware from the constraint of physically replicating the dimensions of the demanded MVMs. In comparison to other cutting-edge photonic CNN architectures evaluated on the MNIST handwritten digits task [17], the OSS accelerator has shown equivalent precision while achieving unmatched efficiency in terms of power consumption and computational density [16]. Lastly, in comparison to a parameter-compact digital CNN, while the OSS accelerator did exhibit slightly lower precision, it managed to reduce the number of floating-point operations by a factor of 50.

In this work, the transition of the OSS acceleration approach into the domain of reprogrammable photonics is explored. Specifically, the photonic front-end of the OSS-CNN system is implemented within a reconfigurable silicon PIC, with the OSS nodes being emulated by physically accurate optical filters, diverging from the ideal Butterworth filters that were explored in previous research. The first section of this document provides an analysis of the operational characteristics and performance relative to the hyperparameters of an ideal OSS-CNN system (no losses, ideal Butterworth filters), in comparison with a PIC-based approach. Numerical simulations conducted using iPronics' Smartlight emulation platform indicate that the proposed reconfigurable photonic OSS accelerator notably can improve performance on benchmark datasets such as MNIST by at least 5.3% in accuracy, when utilized prior to a basic FNN. The subsequent section documents the experimental validation of the OSS accelerator on a reconfigurable silicon photonic chip, achieving a 97.7% accuracy with only 3 OSS nodes while maintaining the same number of data samples at its output as the input visual data. The final section compares the reconfigurable OSS-CNN to a digital equivalent single-layer CNN implemented on a state-of-the-art GPU. The exploration of performance and total power consumption for both architectures is conducted in relation to the depth of the FNN employed at their back-end. Even though the OSS accelerator exhibits slightly inferior precision compared to the single-layer CNN, it significantly reduces the total power consumption during the training stage by an impressive 26%.

## 2. Architecture, Principle of Operation and Simulated OSS System

For In Figure 1, the generic OSS architecture is depicted. Initially, the input tensors are encoded on the intensity of an optical carrier via a Mach-Zehnder Modulator (MZM), effectively encoding the spatial information in the temporal domain. The core of this architecture comprises multiple parallel passive OSS nodes, each consisting of bandpass optical filters (e.g., MRRs, MZIs, etc.). Conceptually, applying the filter's transfer function in the frequency domain is analogous to convolving the signal with the filter's impulse response in the temporal domain. As a result, these proposed nodes can serve as analog convolutional kernels, directly applying distinct sets of complex weights to the incoming time-traces. Unlike conventional CNNs, the weights of these kernels are not arbitrarily controlled but are coarsely tuned through the filter's hyperparameters, which correspond to the parameters forming the impulse response of the OSS filters. For instance, the impulse response of a first-order filter can be expressed as follows:

$$h(t) = 2\pi f_c e^{-t(2\pi f_c - j2\pi f_m)}, \; t > 0 \quad (1)$$

where $f_m$ and $f_c$ correspond to the central and the cut-off frequency of the filter respectively. As shown in Figure 1 (red and green insets), different frequency detunings from the optical carrier and/or different filter bandwidths offer diversified impulse responses in the complex domain (I/Q plots). The receptive field of each kernel, representing the spatio-temporal window in which they are applied, can be adjusted by manipulating the bandwidth and filter order. For instance, wider bandwidth filters lead to a faster exponential decay in the impulse response, thereby encompassing fewer pixels in each convolutional operation. Conversely, the filter's order affects its steepness and introduces a change in the shape of the impulse response, resulting in a slight alteration of the set of applied complex weights.

This generic scheme is accompanied by photodiodes (PDs) and an analog-to-digital converter (ADC) following each node. The digitized signals from the nodes are flattened and fed into a trainable readout layer, implemented in the digital domain. This digital classifier consists of a simple software-based FNN with a softmax activation function at the output layer and an Adam optimizer [18]. Therefore, the proposed scheme pre-processes information directly in the analog domain and it is followed by a conventional yet lightweight digital back-end.

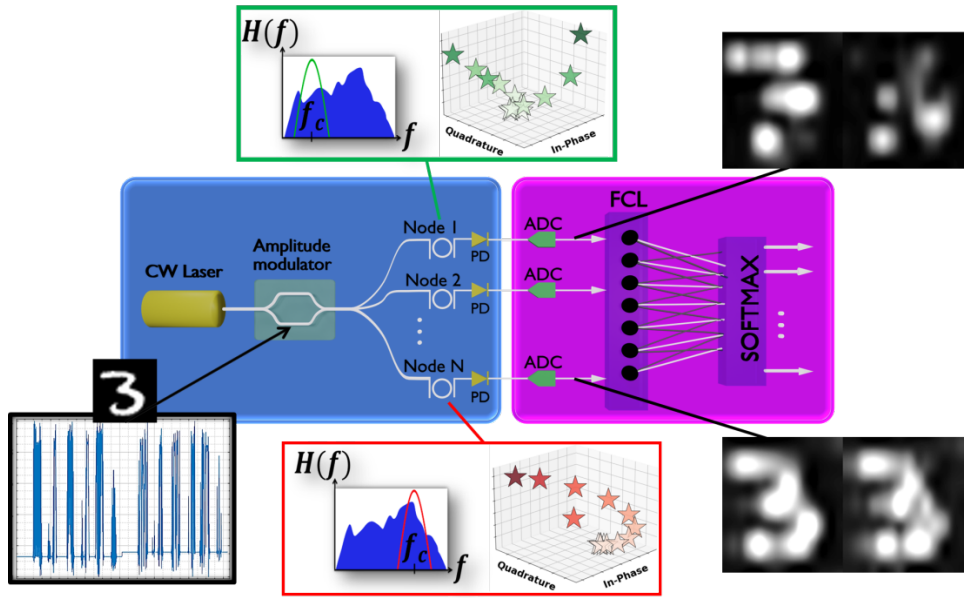

Fig. 1 Schematic diagram of the OSS-CNN architecture. At the insets the different spatial features for the digit "3" are presented by changing the central position of the filter-nodes.

In this study, the acceleration capabilities of the OSS technique were initially examined through simulations. Ideal-generic Butterworth filters, on one hand and a reconfigurable photonic mesh emulator, on the other hand, were used. The primary goal of this investigation was to determine the maximum potential of the approach (ideal system) and then make direct comparisons with real-world implementations, emphasizing on the impact of various physical constraints. For the photonic circuit, we employed the reconfigurable photonic hardware platform (Smartlight), which Software Development Kit enables the incorporation of all the physical non-idealities of a photonic circuit, including factors such as limited resolution during phase tuning, optical coupling errors and random passive phases. Most of these effects are automatically handled and mitigated by the control algorithms. The photonic waveguide mesh is composed of programmable unit cells (PUCs) that enable the implementation of filters using a single MRR in an add/drop configuration (as shown in Fig. 2a). Each MRR-node comprises eight PUCs: one PUC is configured in the cross state to direct the optical signal from the input into the waveguide loop, two PUCs serve as tunable couplers for light input and output from the MRR loop and the remaining five PUCs are set to the bar state to ensure proper light guidance within the loop and at the OSS node's output. Furthermore, each MRR's central frequency can be adjusted by modifying the phase shifter within the optical cavity, allowing each MRR to target a different spectral region of the input, essentially acting as a distinct CNN kernel. Additionally, due to their resonant nature, the MRR filters operate as optical integrators over specific time windows [19], or receptive fields, on the convolved analog responses. Consequently, the output time-traces from the OSS filters involve both the application of weights and their integration, resembling the generation of feature map elements in CNNs.

Chip-related constraints are also applied, including the binding of the OSS nodes' bandwidth, which is linked to the dimensions of the PUCs (823.3 μm length). The node losses are set at approximately 12 dB and the digital-to-analog converters (DACs) responsible for thermo-optic tuning at each PUC operate with a 16-bit resolution. It is important to note that, in the first ideal-node approach, the node losses are disregarded and the bandwidth along with the central frequency of simple optical Butterworth filters are treated as unconstrained hyperparameters.

Finally, thermal and shot noises in the PDs are also taken into account and the ADC was set to an 8-bit resolution at 10 GS/s for both cases.

To assess the scheme's efficiency, the focus is placed on the conventional benchmark task of classifying the handwritten digits of the MNIST dataset consisting of 60,000 images for training and 10,000 for testing. No digital preprocessing has been implemented; instead, a simple pixel-rearranging patching technique has been utilized. Specifically, each image has been partitioned into non-overlapping patches of uniform size ($MxM$) and these patches have been serialized in two orientations (as illustrated in Fig.2b): one along the columns (orientation A) and another along the rows (orientation B). Both orientations have been extracted from each patch and combined into a single vector for each image, enhancing the spatial correlation of their representation in the temporal domain.

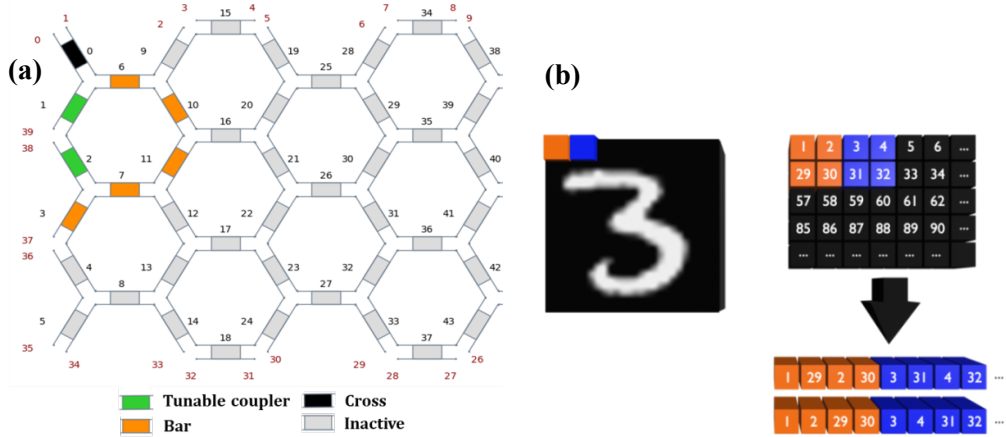

Fig. 2. a) OSS node implementation as a first order MRR at the tunable photonic mesh, b) 2x2 patching of the MNIST images and patch serialization with two orientations: A and B.

Operation-wise, the PDs play a dual role in the system. Firstly, a nonlinear activation function is provided to the optical outputs through their square law behavior, with performance similar to that offered by a ReLU activation [20]. Secondly, they are simulated to be followed by fourth-order Butterworth filters, with a 3 dB bandwidth that is inversely proportional to the used patch size. The bandwidth of the PDs is determined by the formula:

$$BW_{PD} = \frac{PR}{M \cdot M} \quad (2)$$

$PR$ represents the pixel rate (modulation rate at the MZM) of the input signal and $MxM$ the patch size. In this manner, an average value that corresponds to a timeslot of $M^2$ initial pixels is produced at the output of each PD as the optical signal passes through. As a result, this configuration performs both nonlinear activation and average pooling operations in the analog domain, resembling the behavior of a typical CNN. Another degree of freedom is the sampling rate ($SR$) at the ADCs that can be adjusted to control the level of data compression at the input of the FNN. The compression ratio (CR) serves as a useful metric to demonstrate parameter reduction and is defined as the ratio between the size of the input tensor and the FNN's input size. It is expressed as:

$$CR = \frac{H \cdot W \cdot C}{N_{nodes} \cdot N_s} \quad (3)$$

where $H, W$ and $C$ correspond to the height, the width and the number of channels of the input tensor, respectively (28×28×1 for MNIST images), $N_{nodes}$ to the number of utilized OSS nodes and $N_s$ to the number of digital samples at the output of each OSS node. For instance, a $CR =$

1 corresponds to a total of 784 digital samples delivered by the OSS nodes at the input of the FNN, thus indicating zero data compression.

In the ideal node case, the effect of the patching size ($MxM$), the number of OSS nodes ($N$) and the OSS filter characteristics ($f_c, f_m$) were investigated alongside the total input power and the *SR* of the ADCs. Our goal was to identify under which conditions classification accuracy, parameter minimization, throughput and power consumption can be optimized. The bandwidth and the central frequencies of the bandpass filters were initially set as to uniformly cover the input's signal single-sideband spectrum depending on the number of OSS nodes, which were explored for $N = 2$ up to 20. Assuming for example a state-of-the-art 128 GS/s DAC, an intensity modulation setup and a 5 node OSS, the filters cut-off and central frequencies are respectively set to $f_c = 6.4$ GHz and $f_m = 6.4, 19.2, 32, 44.8, 57.6$ GHz. Figure 3 illustrates the classification accuracy as a function of the CR for different numbers of filter nodes on an ideal OSS scheme with a single fully connected layer (FCL) as its digital classifier. The pixel serialization is based on a 4x4 image patching which yielded the best results in terms of accuracy for the MNIST task across all cases. The mean input power was maintained at 20 mW, corresponding to a minimum 53 dB electrical signal-to-noise ratio (SNR) at the receiver of each node, with no observed benefits from further increases due to the absence of losses.

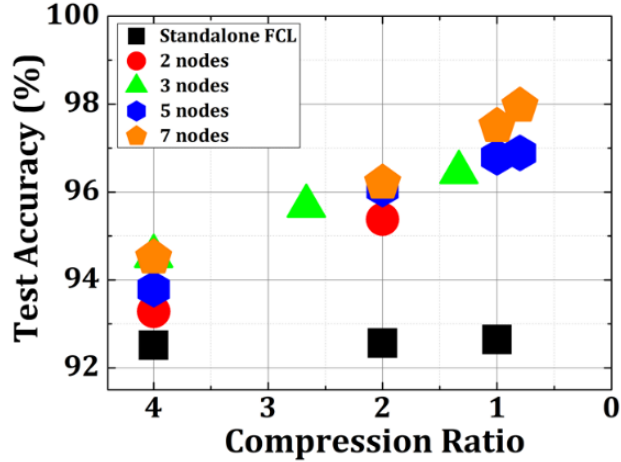

Fig. 3. Classification accuracy of OSS-CNN featuring a single FCL back-end, depicted as a function of the CR and the number of OSS nodes utilized. The analysis is conducted without accounting for chip losses, employing a modulation rate of 128 GS/s and a patch size of 4x4.

As a baseline, feeding all MNIST image pixels directly into a standalone FCL with a softmax activation resulted in an accuracy of 92.6% (Fig. 3 – black points). The classification accuracy of the OSS-CNN is observed to scale inversely with the CR and the beneficial impact of employing more OSS nodes becomes more evident at smaller compressions. The highest accuracy with a single FCL back-end, approximately 98%, was achieved using 7 OSS filters (Fig. 3 – yellow points) and augmenting the number of nodes beyond this did not result in further improvements. Notably, even with increased CR, only a mild performance degradation is observed for the OSS scheme. For instance, at a CR of 4, the performance exceeds that of the standalone FCL, reaching a maximum of 94.6% with 3 and 7 OSS nodes (refer to Fig. 3), compared to the 92.5% achieved by the standalone FCL at this level of compression. It should be noted here that for the baseline case (no OSS preprocessing), the different CR scenarios were obtained by averaging the MNIST image values over varying pixel windows. Specifically, a CR of 4 corresponds to division and averaging across a 2x2 pixel window, while a CR of 2 entails averaging between two consecutive pixels. Therefore, it becomes apparent that the utilization of OSS prior to the FCL consistently enhances classification accuracy across all cases, with the OSS scheme surpassing the maximum accuracy of the standalone readout even with a limited number of nodes and significant compression. This improvement can be attributed to the efficient extraction of differentiated spatio-temporal features from each image by the photonic front-end, even with coarse tuning of the filters' response. Consequently, the OSS preprocessing not only facilitates an enhancement in classification accuracy for the FCL but also provides a valuable dimensionality reduction.

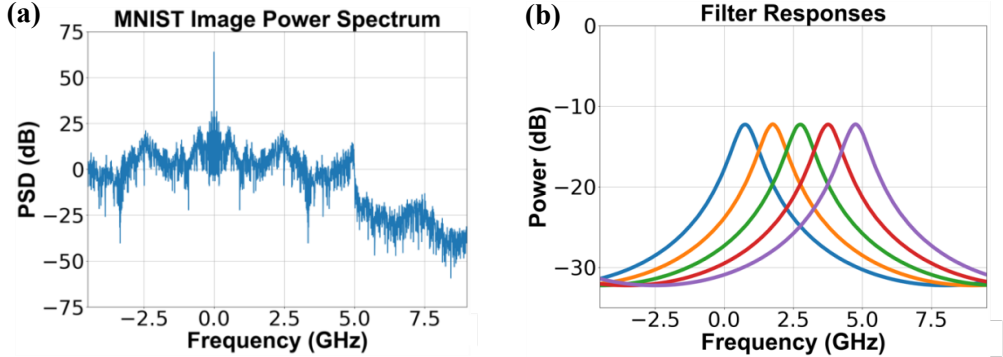

Fig. 4. a) Optical power spectrum of an intensity modulated optical carrier centered at 1550nm with a MNIST image and a 10 GS/s pixel rate, b) Power responses from the drop ports of five Smartlight-implemented OSS nodes, set as to uniformly "slice" the single side-band of the optical spectrum.

As the next step, the Smartlight emulator was employed to implement the OSS scheme. The data serialization, signal detection and digital processing steps remained the same as described above. The configuration of the add/drop MRR filter-nodes utilized is presented in Fig. 2a, featuring a total of 8 PUCs. Specifically, one PUC was configured in the cross state (Fig. 2a - black inset) to guide light from the input port into the MRR loop. Additionally, two PUCs were employed as tunable couplers with coupling coefficients set to $k_1 = k_2 = 0.05$ (Fig. 2a - green) as to couple the light in and out of the loop. The remaining five PUCs were configured in the bar state to guide light within the loop and towards the output port of the OSS node (Fig. 2a - orange). To achieve the appropriate detunings from the optical carrier for each OSS node, the central frequency of each MRR is adjusted by modifying a phase shifter within the loop in order to process a different region of the input signal spectrum. Configuring each OSS node in the waveguide mesh required a total of nine thermal actuators: two for controlling the node's central frequency and one for each of the seven other PUCs within the node. Figure 4 illustrates the power spectrum of an intensity-modulated with a MNIST image optical signal, along with the power responses of the drop ports from five OSS nodes implemented using the Smartlight emulator. These OSS nodes are configured to uniformly 'slice' the single sideband of the signal's spectrum. It can be observed in Figure 4b that setting the insertion losses for each PUC to 0.2±0.02 dB leads to an approximate total loss of 12 dB at the drop port of each OSS node. The input pixel rate was set to 10 GS/s to match the fixed 10 GHz free-spectral range of the photonic mesh filters. This adjustment ensures that each filter interacts with a single region of the optical spectrum exclusively. Lastly, it should be noted that the bandwidth of the mesh-implemented OSS nodes is not an unrestricted hyperparameter. Due to the fixed length of the PUCs, the mesh filters have a fixed 3 dB bandwidth, which is approximately 0.9 GHz.

Figure 5 illustrates the relationship between mean input optical power and classification accuracy for a 5-node OSS implemented on the waveguide photonic mesh, using a 4x4 patch serialization. The comparison is drawn against the maximum accuracies represented by black dashed lines for a standalone FCL and by blue dashed lines for an ideal 5-node OSS setup (no losses, optimum bandwidth and an SNR exceeding 53 dB for each node). The peak accuracy recorded for the PIC-based OSS, at an input power of 20 dBm, stands at 96.7%, reaching an equivalent performance to the ideal OSS-CNN (96.9% with 5 optimized nodes). This slight variance can be attributed to the fixed bandwidth of the PIC filters, which deviates from the optimal bandwidth for processing the MNIST dataset. The analysis also reveals that the OSS-CNN system demands a minimum of 16 dBm power to align closely with the ideal OSS's accuracy. Falling below this threshold leads to performance drops, attributed to the cumulative losses in nodes which increase the noise susceptibility of their outputs. Consequently, the digital signals produced by nodes with lower SNRs become less distinct across different images, complicating the classification task for the FCL. Lastly, it should be noted that absent noise constraints, the impacts of limited hyperparameter tuning resolution (for instance, a 16-bit DAC for phase adjustment) and the non-variable bandwidth on the system's efficacy are minimal.

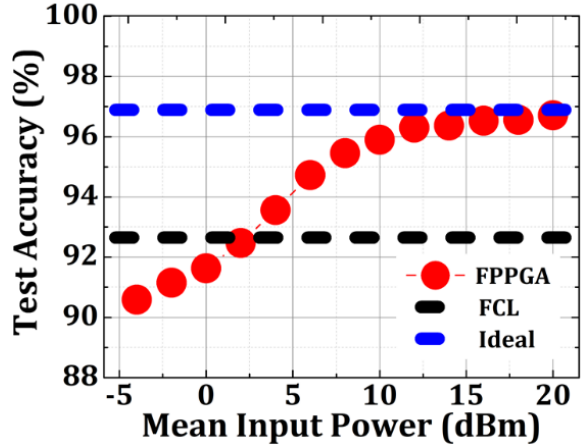

Fig. 5. Classification accuracy of a Smartlight-implemented 5-node OSS-CNN in relation to the mean input power, utilizing a 4x4 image patching technique and a 10 GS/s pixel rate. For comparison, the maximum accuracy achieved by a standalone FCL and that of an ideal 5-node OSS-CNN configuration are also shown.

### 3. Experimental Validation

An exploration of the OSS technique for the MNIST task was conducted through an experiment involving a reconfigurable PIC. In the setup, shown in Fig. 6, a laser emitting light at 1550 nm with an input power of 13 dBm is connected to an intensity modulator. The modulator receives a signal from a 12 GS/s AWG containing the serialized MNIST images, thus enabling a maximum processing speed of approximately 7.6 MImages/s. The serialization of the images follows a 4x4 patch arrangement with a stride of 4, incorporating both patch orientations (see Fig. 2b). Consequently, each image, consisting of 784 pixels, was represented again by a vector of 2×784=1568 samples.

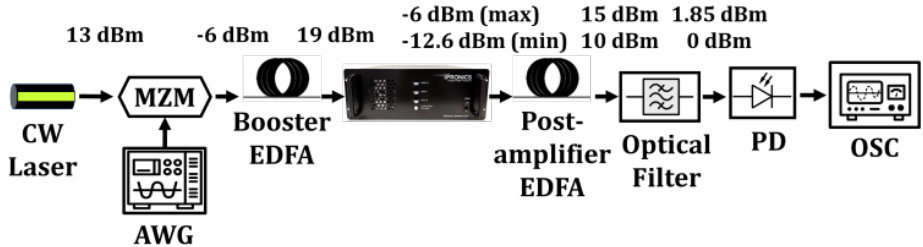

Fig. 6. The experimental setup of the OSS scheme.

The modulated optical signal was subsequently directed to a booster EDFA, where its power level is enhanced to 19 dBm. The optical output of the EDFA was fed into the iPronics Smartlight system [21], a multipurpose optical processor. Within the photonic chip, an MRR filter, as illustrated in Fig. 2a, is implemented to function as the designated OSS node. The resonator's two coupling coefficients were set at 0.1. The power transfer function of the filter for the MRR's drop port, utilized as the output of the OSS node, is shown in Fig. 7. The MRR filter's 3-dB bandwidth is approximately 1.65 GHz, enabling the "slicing" of the 6 GHz single-side bandwidth of the input optical signal. Frequency shifts in the MRR cavity, enabled by a phase shifter, allowed the filter to interact with different spectral components of the input signal. The experiment is conducted in a sequential manner, with the MRR's transfer function undergoing step-by-step frequency shifts and its output recorded for each spectral position. Specifically, the frequency detunings between the carrier frequency and the output range from 0 to 7 GHz, in 0.5 GHz steps, effectively monitoring 15 distinct filter positions.

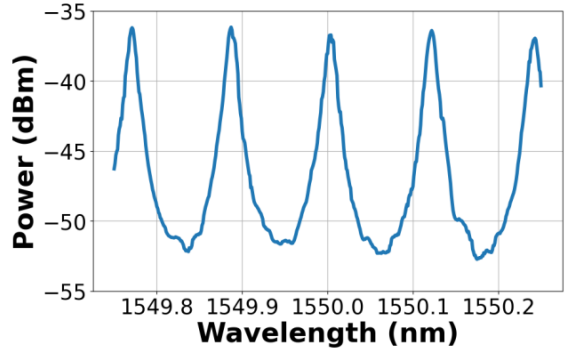

Fig. 7. Power transfer function of the MRR's drop port as obtained from the iPronics Smartlight processor, plotted against wavelength.

Subsequently, the output from the drop port is directed to a second EDFA, which serves to elevate the power level of the optical signal, ensuring effective photodetection by a 20 GHz PD. To reduce the noise from the EDFA, a bandpass filter is positioned before the photodetector. The power budget, illustrating the maximum and minimum transmissivity power levels of the drop port, is also outlined in Fig. 6. Finally, the signal is captured with a 25 GS/s oscilloscope and subsequently resampled at a rate of 12 GS/s, aligning with the size of the initial image-vectors.

After resampling, the image samples are processed through a 4th order Butterworth filter, following the same approach as simulations for the temporal average pooling. In this context, equation 3 can be simplified to $CR = N/(2N_f)$, where $N$ is the number of successive samples that are subjected to the average pooling and $N_f$ represents the number of distinct filter positions used. The factor of 2 in the denominator corresponds to the number of patch representations employed. Following these procedures, the data is reassembled into a $70,000 x K$ matrix, where $K$ denotes the number of digital samples per image at the output of the OSS filters and is given by $K = 784/CR$. The matrix is then divided to create a classification dataset comprising 60,000 digitized outputs for training and 10,000 for testing. A single FCL is used as the system's back-end classifier, with its input size varying according to $K$ and its output size corresponding to the 10 different MNIST classes. Finally, a softmax activation follows the FCL to convert its outputs into probabilities. Fig. 8 provides a schematic representation of the post-processing procedure, indicatively illustrating the process for a single MNIST image for a CR of 2.

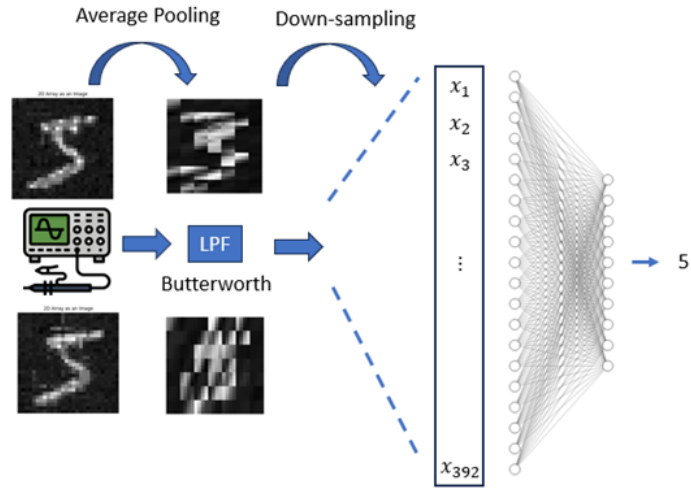

Fig. 8. Illustration of post-processing and classification procedures: A serialized MNIST image undergoes low-pass filtering and is subsequently downsampled by a CR of 2 (K=392 digital samples), preparing it for input into a single FCL.

The classification process included the analysis of combinations of outputs from 2, 3 and 5 distinct filter positions to discover the maximum accuracy achievable by the OSS system. To identify the baseline accuracy, modifications were made to the experimental setup by omitting the photonic processor and the second EDFA. As a result, the modulated signal passed through the first EDFA, the bandpass filter and the PD and was resampled at 12 GS/s after the oscilloscope. The digitized time-traces were subsequently normalized and fed directly into the FCL, which underwent training with an Adam optimizer and a learning rate set at 0.01. Figure 9 illustrates the peak classification accuracies attained by the experimental OSS system, correlating them with the number of distinct filter position outputs fed into the FCL classifier. Analyzing the outputs from two filter positions yielded a maximum accuracy of approximately 95.7% at a CR of 1, achieved by combining the filter at the carrier frequency (0 GHz detuning) with a filter detuned by 2.5 GHz. The utilization of three filter positions, specifically at 0, 2.5 and 7 GHz detunings, produced the most advantageous results, achieving a top accuracy of 95.87% with a CR of 1. It should be emphasized that the precision achieved by this experimental 3-node OSS exceeded the highest experimental performance reported for state-of-the-art photonic CNN accelerators, such as [13,14]. Lastly, incorporating outputs from five distinct filter positions did not further enhance accuracy, peaking at 95.76% with the inclusion of five filter positions.

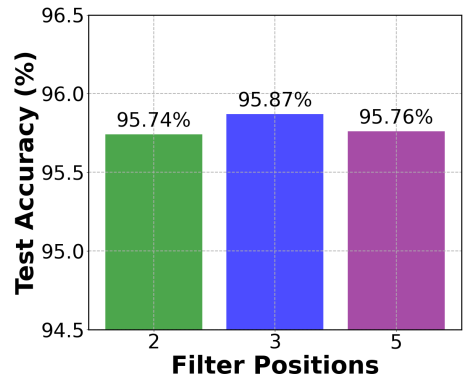

Fig. 9. Maximum accuracy of the experimental OSS-CNN as a function of the number of distinct filter position outputs employed as inputs on a single FCL classifier.

The classification accuracy of the experimental OSS system with a 1-FCL at its back-end, for the three most effective filter positions in relation to the CR is depicted in Figure 10. This is compared against the baseline accuracy, as indicated by the black dashed line in the figure. To provide context for these experimental findings, two scenarios featuring simulations of ideal OSS nodes are also incorporated in the figure. The first scenario demonstrates a generic solution in which the central frequencies of the nodes are fixed to uniformly cover the single sideband spectrum of the optical signal, as shown by the blue plot in Figure 10. The second scenario examines a configuration in which the central frequencies of the filter nodes are considered as free hyperparameters for the system. In more detail, an investigation was executed with the aid of the 'Optuna' hyperparameter optimization framework [22], aiming to maximize the precision of an ideal 3-node OSS-CNN. The optimization task in this framework is usually defined by specifying the hyperparameter search space and an objective function to minimize or maximize. Additionally, advanced algorithms such as the Tree-structured Parzen Estimator (TPE) and CMA-ES are utilized to intelligently navigate the search space, offering a more efficient solution than traditional methods such as random or grid search. In this work, the bandwidth and central frequencies of the OSS filters were examined as hyperparameters, with the latter being able to take any value within the single side-band spectrum of the input signal. A TPE algorithm was employed to predict promising hyperparameter combinations based on past trial outcomes that would maximize the classification precision which was set as the objective function. The patch size was set to 4x4, the input power at 20 mW and the sampling rate was adjusted to extract two samples per patch. The optimum setup emerged for $f_m$ values of 6, 22 and 27 GHz and $f_c$ equal to 5 GHz and is depicted by the green plot in Figure 10. In both ideal OSS scenarios, the method employed for extracting digitized sequences from each OSS node was the same as the experimental post-processing procedure. The OSS node outputs were processed through a simulated 20 GHz PD and a digital 4th order Butterworth filter, adjusting its bandwidth to match the digital sample count at the FCL input, as per the experimental setup.

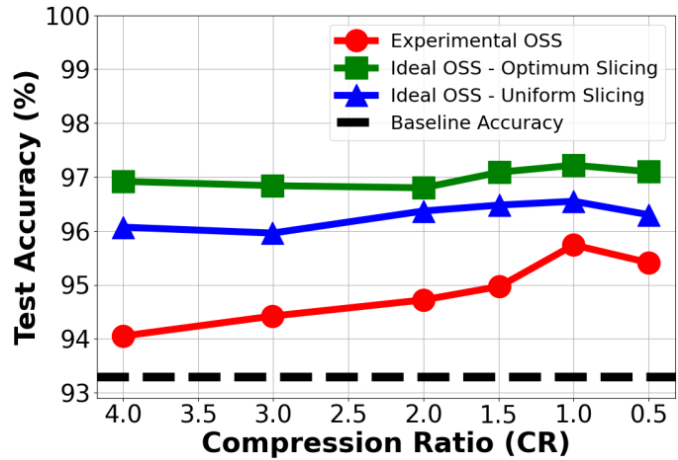

Fig. 10. Classification accuracy of the experimental OSS system with a 1-FCL back-end across the most effective filter positions (0, 2.5 and 7 GHz) in relation to the CR. Included are the baseline accuracy (black dashed line) and two scenarios of an ideal 3-node OSS-CNN: one with central frequencies fixed to uniformly cover the optical signal's single sideband spectrum (blue plot) and another where the central frequencies are set at the optimum positions for the MNIST task (green plot).

The capabilities as well as the robustness of the OSS preprocessing are highlighted in Figure 10. First, it is observed that the classification accuracy of the experimental setup consistently surpasses the baseline accuracy, particularly in cases with lower CRs. The peak accuracy, observed at 95.9% with a 1-FCL classifier, surpasses the baseline system accuracy by 4.3%. This notable performance improvement not only confirms the efficiency of the Smartlight OSS as an accelerator but also outperforms the experimental results of other state-of-the-art photonic schemes, as referenced in [13,14]. Secondly, it is evident that the experimental OSS manages to maintain a higher level of accuracy compared to a standalone FCL fed with the entire MNIST image data, even in scenarios involving significant data compression/averaging. For example, the experimental OSS-CNN achieved an accuracy of 94.1% with a CR of 4 and approximately 94.7% with a CR of 2, surpassing the baseline by 1.5% and 2.1% respectively. Therefore, even with the presence of losses, the amplified spontaneous emission introduced by the EDFA and the fixed bandwidth of the Smartlight-implemented nodes, the OSS was able to deliver efficient parameter reduction. Thirdly, when comparing the experimental OSS to the ideal OSS-CNN models, the experimental OSS's accuracy demonstrated a trend similar to the ideal cases, inversely scaling with the CR up to a CR of 1, beyond which a slight decrease in accuracy was noted for CR=0.5. A noticeable difference between the ideal and experimental setups was seen in the rate at which accuracy decreased as the CR increased. This decline in the experimental OSS's performance can be attributed to two main reasons. Firstly, the bandwidth of the filters implemented in the Smartlight processor deviates from the optimal bandwidth required for this specific task. Secondly, considering the vast array of possible combinations of three filters from all 15 distinct filter position outputs, our investigation was limited to only those combinations that included the filter positions yielding the best outcomes in the 2-node scenario (0 and 2.5 GHz detunings). Lastly, comparing the ideal OSS-CNN configurations, it is observed that the performance of the generic OSS setup, where the central frequencies of the nodes uniformly filter the single-sideband spectrum of the input signal, is slightly to that of the OSS setup with optimized filter positions tailored for the specific task. For instance, in the optimized OSS scenario, the peak accuracy attained was 97.2% at a CR of 1, which only slightly decreases to 96.9% at a CR of 4. Conversely, in the generic OSS scenario, the highest observed accuracy was 96.5% at a CR of 1, dropping to 95.9% at a CR of 4. This depicts the OSS-CNN's ability to achieve sufficient precision in machine vision tasks by uniformly covering the single side-band of the input spectrum, without the need to know the optimum filter positions. It should be noted that, with sufficient input power, comparable performance was also achieved in numerical simulations realizing 3-node OSS-CNNs in the Smartlight emulator.

## 4. Performance Comparison with a Respective Digital CNN

In this section, a performance evaluation compares the OSS accelerator with a conventional digital CNN architecture, which features a single convolutional-nonlinear-pooling stage. For the classification back-end, two scenarios are depicted: a single FCL equipped with a softmax activation function and configurations including an additional hidden FCL, while adding a second hidden layer did not enhance performance in any case. Table 1 presents a detailed comparison of key characteristics and metrics for three architectures: the optimally simulated 7-node OSS accelerator, a corresponding single-layer digital CNN utilizing the same number of kernel filters, and the optimal 3-node experimental OSS. The digital back-ends of the OSS schemes, as well as the entire single-layer CNN, were implemented on the TensorFlow platform [23] using an NVIDIA GeForce RTX 2080 Ti GPU [24]. To obtain the maximum classification accuracy for each scheme, an investigation was conducted using the 'Optuna' platform with respect to the number of nodes per layer, learning rate, batch size, and the activation function of the hidden layer. In the case of the single-layer CNN, the size of the kernels and the pooling were also investigated. Table I details the number of hidden FCLs, units per layer, total trainable parameters and the classification accuracy for the optimal setup of each architecture. Additionally, the average power consumption of the graphics card ($\bar{P}_{GPU}$) during the training procedure is depicted for each architecture. These values were derived from monitoring the mean Thermal Design Power percentage (TDP%) during the training process for each approach. In general, TDP% is a metric that indicates the GPU's power usage and thermal load. Given that the TDP of the GPU used in this study is 250 W, the TDP percentage was calculated using the GPU-Z monitoring utility [25]. The average TDP percentage was computed over 100 training epochs with a consistent learning rate and batch size for both architectures.

**Table 1. Performance Comparison between the OSS Accelerator and a Single-Layer Digital CNN**

| Scheme | Number of FCLs (units per layer) | Trainable Parameters | MNIST Accuracy (%) | $\bar{P}_{GPU}$ (W) |
|---|---|---|---|---|
| Single-Layer CNN | 1 (10) | 8,830 | 98.6 | 50 |
|  | 2 (117,10) | 41,778 | 99.1 | 50 |
| 7-node OSS-CNN | 1 (10) | 13,730 | 98 | 33.3 |
|  | 2 (111,10) | 153,523 | 98.6 | 34.3 |
| 3-node Experimental OSS-CNN | 1 (10) | 15,690 | 95.9 | 32.8 |
|  | 2 (80,10) | 184,753 | 97.7 | 34.3 |

The information from Table 1 reveals that adding a second FCL at the back-end, the simulated OSS exhibited a maximum accuracy of 98.6%, marginally lower than the performance of the respective single-layer CNN, which scored 99.1%. This same difference in precision is also observed with one dense layer, with the optimized digital CNN recording a 98.6%. Regarding the experimental OSS, it is clear that the introduction of an additional FCL boosted its accuracy from 95.9% to 97.7%, thus narrowing its performance gap to under 1% compared to the simulated OSS, which utilized more nodes, and to only 1.4% compared to a fully optimized noiseless digital system. However, the OSS-CNN displayed a notable advantage in energy efficiency. Specifically, both for the simulated and experimentally produced inputs for the digital back-end, the OSS-CNN managed to reduce the average power consumption of the graphics card by a significant 33% for configurations with one and two FCLs. This reduction in power consumption is particularly noteworthy because the OSS scheme, despite having a higher number of trainable parameters compared to the digital CNN across all scenarios, manages to be more energy-efficient due to the opto-electronic pre-processor.

This pre-processor effectively decouples the convolutional, nonlinear, and pooling stages from the digital domain, significantly reducing the computational workload of the GPU, which must handle all stages in the single-layer CNN. For instance, in a traditional setup, each of the 7x7 kernel filters within the convolutional layer of the digital CNN performs convolutions at 22x22 positions across each MNIST image when striding with a step of 1. Therefore, on a single pass, each kernel filter executes $2 \cdot 7 \cdot 7 \cdot 22 \cdot 22 = 47{,}432$ operations for each MNIST image, encompassing both multiplications and additions. Collectively, all seven kernels in the convolutional layer execute a total of 332,024 operations. Additionally, the subsequent single-layer feedforward network with 10 units handles 16,940 operations, derived from 847 multiplications and additions per unit. Thus, the model processes approximately 348,964 operations in one forward pass, not including the operations from the pooling layer. Conversely, in the OSS accelerator, the scope of forward and backward pass computations during the training process is limited to the depth of the digital back-end classifier. For example, in the single FCL case, with 10 units and an input size of 1,372 (total samples per MNIST image at the output of the 7 OSS nodes), a total of $10 \cdot 1{,}372 \cdot 2 = 27{,}440$ operations are executed in a single forward pass. Hence, approximately 92% fewer computations are executed within the

GPU with the OSS preprocessing, not accounting for the backpropagation operations during training, which typically amount to twice the number of forward pass computations, as well as the power consumption from memory accesses and data transfers between memory and GPU [26].

To compare the OSS-CNN with this conventional digital CNN architecture in terms of end-to-end power consumption, it is necessary to consider not only the GPU's power consumption but also that of the opto-electronic pre-processor and the digital post-processor components. The formula for calculating the total power consumption of a Smartlight OSS-CNN can be given by:

$$P_{total} = P_{LD} + P_{OSS} + P_{DAC} + P_{ADCs} + \bar{P}_{GPU} \quad (3)$$

where $P_{LD}$ is the power consumption of the laser source, $P_{OSS}$ is the total power consumption to implement the OSS nodes in the waveguide mesh, $P_{DAC}$ is the power of the front-end DAC and $P_{ADCs}$ is the total power draw of the ADCs. Based on [27], the power consumption of a laser diode is given by:

$$I_{LD,driv} = \frac{P_{LD,opt}}{\gamma_{LD}} + I_{LD,thr} \quad (4)$$
$$P_{LD} = V_{LD,f} \cdot I_{LD,driv} \quad (5)$$

where $V_{LD,f}$ is the diode bias voltage, $I_{LD,driv}$ is the diode current, $P_{LD,opt}$ is the optical power of the laser, $\gamma_{LD}$ is the slope efficiency and $I_{LD,thr}$ is the threshold current. For a typical continuous-wave 1550 nm laser diode emitting at 16 dBm [28], the values are as follows: $V_{LD,f} = 1.8$ V, $I_{LD,thr} = 8$ mA and $\gamma_{LD} = 0.29$ mW/mA. With a typical continuous-wave 1550 nm laser diode emitting at 16 dBm, the estimated power consumption is approximately 262 mW. Similarly, the power consumption for a state-of-the-art front-end DAC operating at 50 GS/s is about 243 mW [29]. The power consumption for the photonic waveguide mesh implementing the 7-node OSS scheme is estimated by considering two components for the nine thermal actuators associated with each MRR node: 1.3 mW for maintaining the PUC's state [30] and 33 mW for the DACs driving the actuators [31]. Therefore, the total power consumption for all seven OSS nodes in the mesh (7x9=63 actuators) is approximately 2.18 W. Moreover, the power consumption of seven state-of-the-art ADCs [32], following the number of OSS nodes, that operate at 6 GS/s (in alignment with equation 2 for the 4x4 patching), is calculated to be 105.7 mW. Finally, according to equation 3, the total power consumption of the OSS accelerator can be as high as 37.09 W. This represents a significant reduction in total power consumption, by at least 26%, compared to the respective single-layer CNN, which required an average of 50 W of GPU power for training on the MNIST task.

Finally, a comparison in terms of training time between the OSS and the digital CNN was also performed. In specific, ten iterations of the training procedure were executed for each architecture to calculate their mean training time and the average number of epochs required for convergence, maintaining the same training parameters but varying the initial weight values. For the OSS with a 1-FCL back-end, the average training time per iteration was approximately 224 seconds, with an average of 754 epochs needed for convergence. This contrasts markedly with the single-layer CNN with a 1-FCL back-end, which required only about 17 seconds per iteration and 60 epochs on average. This difference can be attributed to noise introduced by the PDs in the OSS, as well as to representation overheads at the output of the OSS nodes caused by phase-shifting or beam-splitting errors during photonic processing and by the analog-to-digital conversion. With a more complex 2-FCL classifier at its back-end, the OSS scheme demonstrated notably improved efficiency, taking on average approximately 28 seconds per iteration and 96 epochs, reducing the difference to the single-layer CNN with a 2-FCL back-end to less than 4 seconds in training time and 10 epochs for convergence (24 seconds per iteration and 86 epochs on average). It is evident that by adding a second dense layer, the classifier was able to converge significantly faster, comparably to the noiseless digital CNN, due to its increased capacity to learn more complex representations from data that included noise and variations introduced by the OSS processing stages.

## 5. Conclusion

In summary, this work proposes an analogue accelerator based on an integrated photonic reconfigurable platform, circumventing the issues that arise in hardware architectures replicating MVMs. The proposed hardware-friendly technique allows convolutional operations to be carried out by passive optical filter-nodes without requiring fine control or elaborate training, while other essential CNN operations, such as nonlinear transformation and pooling, are also executed in the analog domain without any processing latency. Numerical results confirmed the scheme's ability to achieve a classification accuracy of 98.6%, closely matching performance of full-scale digital networks. Experimental validation of the OSS concept using iPronic's Smartlight reconfigurable photonic mesh offered an accuracy of 97.7%, thus outperforming state-of-the-art photonic implementations, on the MNIST dataset. More importantly the proposed scheme confirms its accelerator properties by drastically reducing the total power consumption by at least 26% compared to its respective standalone digital neural networks. The experimental performance and the reduced power consumption is substantiating the role of photonic accelerators as promising sub-systems for large scale hybrid machine learning schemes.

**Funding.** This work was funded by the EU H2020 NEoteRIC project (871330) and by the EU Horizon Europe PROMETHEUS project (101070195).

**Acknowledgments.** The authors would like to sincerely thank Dr. Daniel Perez Lopez for the fruitful discussions.

**Disclosures.** The authors declare no conflicts of interest.

**Data availability.** Codes for the numerical simulations are available upon reasonable request to the authors.

## References

1. M. H. Miraz, M. Ali, P. S. Excell, and R. Picking, "A review on Internet of Things (IoT), Internet of Everything (IoE) and Internet of Nano Things (IoNT)," in *2015 Internet Technologies and Applications (ITA)* (2015), pp. 219–224.
2. J. Backus, "Can programming be liberated from the von Neumann style? a functional style and its algebra of programs," Commun. ACM **21**, 613–641 (1978).
3. A. Mehonic and A. J. Kenyon, "Brain-inspired computing needs a master plan," Nature **604**, 255–260 (2022).
4. Y. Shen, N. C. Harris, S. Skirlo, M. Prabhu, T. Baehr-Jones, M. Hochberg, X. Sun, S. Zhao, H. Larochelle, D. Englund, and M. Soljačić, "Deep learning with coherent nanophotonic circuits," Nature Photonics **11**, 441–446 (2017).
5. G. Tanaka, T. Yamane, J. B. Héroux, R. Nakane, N. Kanazawa, S. Takeda, H. Numata, D. Nakano, and A. Hirose, "Recent advances in physical reservoir computing: A review," Neural Networks **115**, 100–123 (2019).
6. D. Pérez, I. Gasulla, P. Das Mahapatra, and J. Capmany, "Principles, fundamentals, and applications of programmable integrated photonics," Adv. Opt. Photon. **12**, 709 (2020).
7. Y. LeCun, Y. Bengio, and G. Hinton, "Deep learning," Nature **521**, 436–444 (2015).
8. "Minimizing Computation in Convolutional Neural Networks | SpringerLink," https://link.springer.com/chapter/10.1007/978-3-319-11179-7_36.
9. S. Chetlur, C. Woolley, P. Vandermersch, J. Cohen, J. Tran, B. Catanzaro, and E. Shelhamer, "cuDNN: Efficient Primitives for Deep Learning," (2014).
10. P. Yao, H. Wu, B. Gao, J. Tang, Q. Zhang, W. Zhang, J. J. Yang, and H. Qian, "Fully hardware-implemented memristor convolutional neural network," Nature **577**, 641–646 (2020).
11. H. Zhou, J. Dong, J. Cheng, W. Dong, C. Huang, Y. Shen, Q. Zhang, M. Gu, C. Qian, H. Chen, Z. Ruan, and X. Zhang, "Photonic matrix multiplication lights up photonic accelerator and beyond," Light Sci Appl **11**, 30 (2022).

12. V. Bangari, B. A. Marquez, H. Miller, A. N. Tait, M. A. Nahmias, T. F. de Lima, H.-T. Peng, P. R. Prucnal, and B. J. Shastri, "Digital Electronics and Analog Photonics for Convolutional Neural Networks (DEAP-CNNs)," IEEE J. Select. Topics Quantum Electron. **26**, 1–13 (2020).
13. J. Feldmann, N. Youngblood, M. Karpov, H. Gehring, X. Li, M. Stappers, M. Le Gallo, X. Fu, A. Lukashchuk, A. S. Raja, J. Liu, C. D. Wright, A. Sebastian, T. J. Kippenberg, W. H. P. Pernice, and H. Bhaskaran, "Parallel convolutional processing using an integrated photonic tensor core," Nature **589**, 52–58 (2021).
14. X. Xu, M. Tan, B. Corcoran, J. Wu, A. Boes, T. G. Nguyen, S. T. Chu, B. E. Little, D. G. Hicks, R. Morandotti, A. Mitchell, and D. J. Moss, "11 TOPS photonic convolutional accelerator for optical neural networks," Nature **589**, 44–51 (2021).
15. H. Bagherian, S. Skirlo, Y. Shen, H. Meng, V. Ceperic, and M. Soljacic, "On-Chip Optical Convolutional Neural Networks," (2018).
16. A. Tsirigotis, G. Sarantoglou, M. Skontranis, S. Deligiannidis, K. Sozos, G. Tsilikas, D. Dermanis, A. Bogris, and C. Mesaritakis, "Unconventional Integrated Photonic Accelerators for High-Throughput Convolutional Neural Networks," Intelligent Computing **2**, 0032 (2023).
17. "MNIST handwritten digit database, Yann LeCun, Corinna Cortes and Chris Burges," http://yann.lecun.com/exdb/mnist/.
18. D. P. Kingma and J. Ba, "Adam: A Method for Stochastic Optimization," (2017).
19. M. Ferrera, Y. Park, L. Razzari, B. E. Little, S. T. Chu, R. Morandotti, D. J. Moss, and J. Azaña, "On-chip CMOS-compatible all-optical integrator," Nat Commun **1**, 29 (2010).
20. S. Chen, X. Wang, C. Chen, Y. Lu, X. Zhang, and L. Wen, "DeepSquare: Boosting the Learning Power of Deep Convolutional Neural Networks with Elementwise Square Operators," (2019).
21. "Product," iPronics Programmable Photonics (n.d.).
22. T. Akiba, S. Sano, T. Yanase, T. Ohta, and M. Koyama, "Optuna: A Next-generation Hyperparameter Optimization Framework," in *Proceedings of the 25th ACM SIGKDD International Conference on Knowledge Discovery & Data Mining*, KDD '19 (Association for Computing Machinery, 2019), pp. 2623–2631.
23. "TensorFlow," https://www.tensorflow.org/.
24. "GeForce RTX 20 Series Graphics Cards and Laptops," https://www.nvidia.com/en-eu/geforce/20-series/.
25. "TechPowerUp," https://www.techpowerup.com/gpuz/.
26. T.-J. Yang, Y.-H. Chen, and V. Sze, "Designing Energy-Efficient Convolutional Neural Networks Using Energy-Aware Pruning," in *2017 IEEE Conference on Computer Vision and Pattern Recognition (CVPR)* (IEEE, 2017), pp. 6071–6079.
27. E. Berikaa, M. S. Alam, S. Bernal, R. Gutiérrez-Castrejón, W. Li, Y. Hu, B. Krueger, F. Pittalà, and D. V. Plant, "Next-Generation O-band Coherent Transmission for 1.6 Tbps 10 km Intra-Datacenter Interconnects," J. Lightwave Technol. 1–10 (2023).
28. "1550 nm laser diode up to 400mW -SHIPS TODAY- fiber DFB - pulse&CW 1550nm laser diode," Aerodiode (n.d.).
29. H. Chandrakumar, T. W. Brown, D. Frolov, Z. Tuli, I. Huang, and S. Rami, "A 48-dB SFDR, 43-dB SNDR, 50-GS/s 9-b 2×-Interleaved Nyquist DAC in Intel 16," IEEE Solid-State Circuits Letters **5**, 239–242 (2022).
30. D. Pérez-López, A. Gutierrez, D. Sánchez, A. López-Hernández, M. Gutierrez, E. Sánchez-Gomáriz, J. Fernández, A. Cruz, A. Quirós, Z. Xie, J. Benitez, N. Bekesi, A. Santomé, D. Pérez-Galacho, P. DasMahapatra, A. Macho, and J. Capmany, "General-purpose programmable photonic processor for advanced radiofrequency applications," Nat Commun **15**, 1563 (2024).
31. "MAX5633 Datasheet and Product Info | Analog Devices," https://www.analog.com/en/products/max5633.html.
32. I.-M. Yi, N. Miura, H. Fukuyama, and H. Nosaka, "A 15.1-mW 6-GS/s 6-bit Single-Channel Flash ADC With Selectively Activated 8× Time-Domain Latch Interpolation," IEEE Journal of Solid-State Circuits **56**, 455–464 (2021).